# Optimal Closed Loop Control of G2V/V2G Action Using Model Predictive Controller

Satya Vikram Pratap Singh
*Deptt. of EECS*
*Indian Institute of Technology*
Bhilai, India
satyavp@iitbhilai.ac.in

Siddharth Kamila
*Deptt. of EECS*
*Indian Institute of Technology*
Bhilai, India
siddharthk@iitbhilai.ac.in

Prashanth Agnihotri
*Deptt. of EECS*
*Indian Institute of Technology*
Bhilai, India
pagnihotri@iitbhilai.ac.in

*Abstract*—This paper has developed a closed-loop control algorithm to operate the G2V/V2G action, tested under varying battery voltage conditions and load and source power differences. Under V2G action, to maintain total harmonic distortion under minimum level and grid frequency under the standard limit, a Model predictive controller (MPC) has been used to control the gate driver circuit of the inverter. The state space model of the plant has been created using the system identification toolbox, and the MPC Controller block has been designed using the Model Predictive Control Toolbox of MATLAB. The proposed methodology is tested using MATLAB/Simulink and OPAL-RT (OP4510) in a real-time environment. This methodology reduces %THD to less than 0.5%, improves waveform quality of grid voltage, inverter output voltage, grid current, and inverter output current to nearly 99%, and maintains the grid frequency in standard limit while in G2V/V2G action.

## I. INTRODUCTION

People are drawn to EVs most for their ability to lower $CO_2$ levels at a minimum level compared to internal combustion vehicles [1]. The essential component of electric vehicles, batteries, account for between 30-40% of the total cost of the vehicle and depend on the power grid and micro-grid for charging purposes [2]. In the growing trend of EVs, its chargers, i.e. unidirectional or bidirectional, must be capable of efficient coordination between grid and network to reduce the burden on electric power system networks [3], [4]. Uni-directional (G2V) chargers are cheaper but have flexibility constraints such as compatibility related to vehicles, charging speed, power availability, and cost; at the same time, bidirectional (G2V/V2G) chargers are more flexible, reduce stress on the power networks, and have the potential to accomplish grid balance without expanding power generation infrastructure [3], [4]. During V2G action, bidirectional chargers also provide cheap and fast decentralised energy storage during the overproduction of electricity and serve as an income source that lowers its operational cost [3], [4].

Different control methods have been used for controlling the power flow in the bidirectional charger, such as the proportional-integral (PI) control, the voltage-oriented control (VOC), hysteresis current control, fuzzy logic control, sliding mode control and direct power control (DPC) [6]–[10], [15], [18]. In the case of PI control and VOC, an external voltage loop and inner current loop is required, and also they need modulation and synchronization [5], [6], [15]. In PI control, there is difficulty in tuning the $K_p$ and $K_i$ parameters of both the outer loop and inner loop to an exact value with the change in the system's operating conditions [6]. The Direct power control (DPC) method doesn't need the phase-locked loop (PLL), internal current loop, outer voltage loop, or modulators [10], but it introduces high power ripples, which leads to highly distorted grid currents [11]. Model predictive controllers (MPC) in bidirectional chargers are an attractive solution to overcome the limitations of state-of-the-art control methods. MPC is a numerical optimization-based control that takes the current state as the initial state of the optimal control problem and obtains the optimal control behaviour through the prediction model and cost function at a specified sampling rate [12]. Though it has much computational burden, fast and powerful microprocessors are now available to realise the MPC to control converters in power electronics applications. [12], [13]. Modern MPC algorithm, in addition to operating EV chargers in G2V and V2G mode (to control active power flow between grid and EV charger), can also be used in V4G mode (as a reactive power compensator) to regulate power quality and voltage compensation [14], [16], [17], [19]. To reduce the burden and maintain good power quality on the primary power grid, most MPC algorithms have been developed to integrate the EV charging system with the primary power grid, microgrid, and battery energy storage system [20], [21]. Direct predictive power control (DPPC) algorithm has been developed to improve the system's stability and reliability, which can directly control active and reactive power flow between the grid and EV charger, thus avoiding using PLL [22], [23].

Researchers have not used any closed-loop control strategy for bidirectional chargers in the above works. During V2G action, the quality of voltage, current and frequency fed to the grid, all together, has not been quantified in single work. During G2V/V2G action, information about the grid's frequency to be in standard limit has also not been depicted. The proposed methodology has the advantage of (a) closed-loop control of bidirectional chargers considering the factors such as load power, source power, and battery voltage. (b)

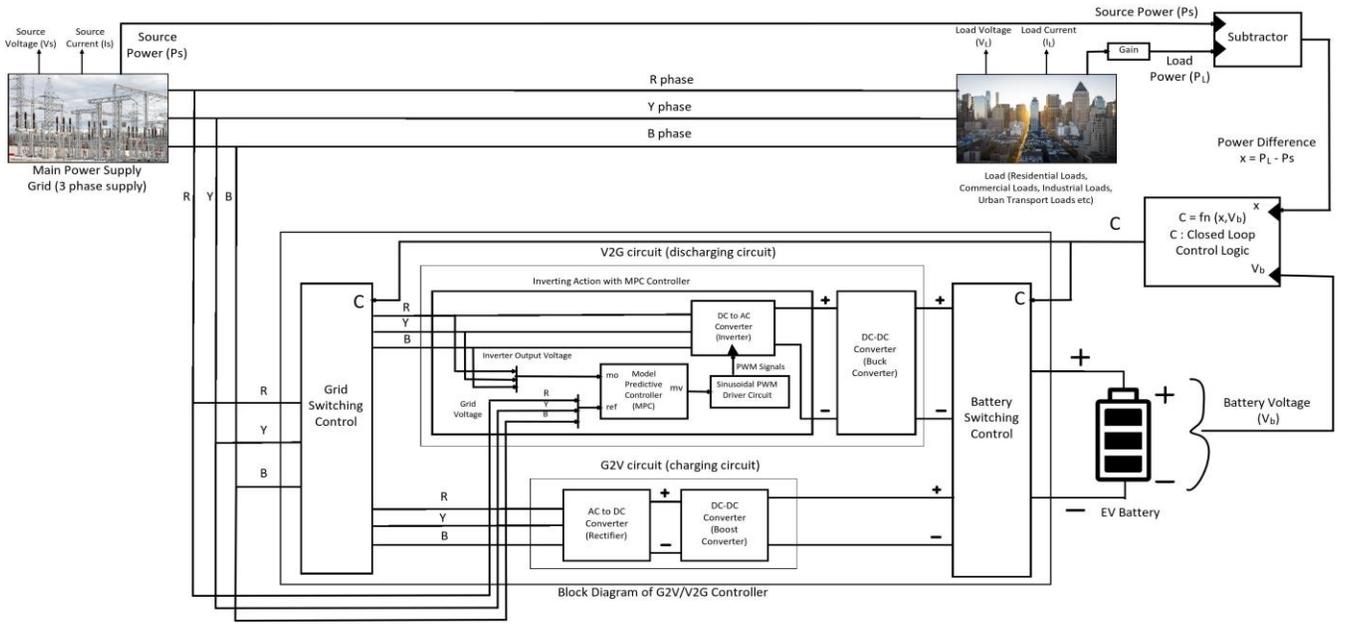

Fig. 1. Block Diagram of G2V/V2G action using MPC controller

Quantifying voltage and current quality during V2G action using the total harmonic distortion factor. (c) Grid's frequency is maintained in the standard limit in both control actions. This paper comprises four sections: Section II describes the proposed methodology, which has a subsection of Block diagram, process flow chart, closed-loop control algorithm, state space modelling of plant, mathematical modelling and design of MPC. Section III highlights the real-time digital environment results, and the paper finally ends with the conclusion in section IV.

## II. Proposed Methodology

In the proposed methodology, considering battery voltage and the difference between source and load power, closed-loop control of G2V/V2G action has been done. In V2G mode, as power is fed back to the grid, the model predictive controller (MPC) has been used to maintain total harmonic distortion (THD) to a minimum level. In MPC, the measured output is the three-phase output voltage of the inverter, and the reference voltage is the three-phase grid voltage.

### A. Block Diagram of G2V/V2G action using MPC controller

The block diagram of G2V/V2G action has been shown in Fig.1. The main 3-phase supply R,Y and B is connected to load in upper path and with EV battery in lower path. In lower path, there are two switching controllers i.e. Grid switching controllers and Battery switching controllers, which is controlled using closed loop control algorithm of Fig.3. In between both controllers, charging and discharging circuits has been shown in Fig.1. In charging circuit, AC supply is converted to DC using three phase controlled rectifier circuit for charging purpose and in discharging circuit, DC supply of battery is converted to AC using three phase inverter circuit.

To maintain %THD of inverter output to an optimal level, the MPC controller has been used to control the gate driver circuit of inverter. As shown in Fig.1, reference is taken as grid voltages and measured output is inverter voltage.

### B. Process flow chart

In the process flow chart shown in Fig. 2, firstly, the value of $x$ defined as $P_L - P_S$, i.e. the difference between load power and source power, and $V_b$, i.e. battery voltage has been calculated. Then these two values are fed to the closed-loop control algorithm, which generates a control logic "c" based on the logic table as shown in Fig.3. Based on "c", G2V/V2G action is being performed. Further, when "c=1", i.e. V2G action (inverting) power is fed back to the grid, in this action, the total harmonic distortion (THD) of voltage and current should be nearly equal to zero, and grid frequency must be in standard limit. Model predictive controller (MPC) has been used to maintain these conditions, which takes three-phase inverter output voltage as measured output (mo) and grid voltage as reference (ref) as shown in Fig.1. Then MPC generates a manipulated variable (mv) signal, which is fed to a sinusoidal PWM generator for switches $SW_n; n \in [1-6]$. The output of the SPWM generator is fed to the gate driver circuit of the inverter.

### C. Closed loop control algorithm

In the closed-loop control algorithm, $x$ and $V_b$ values are considered, as shown in Fig.1. At $x \geq 0$ and $x < 0$, three battery levels have been specified. The upper bound is regarded as $0.75(V_{rated})$, and the lower bound is considered as $0.2(V_{rated})$ in which $V_{rated}$ is the nominal capacity of the battery voltage. To check all the conditions of the closed-loop algorithm, $x < 0$ is created by multiplying arbitrary gain in $P_L$.

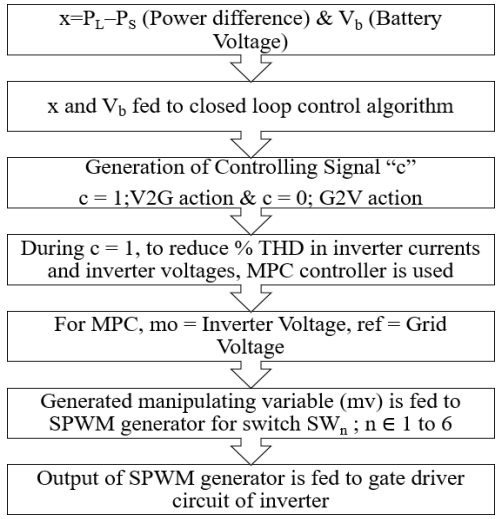

Fig. 2. Process flow chart

| S.No. | x=$P_L$-$P_s$ $P_l$=Load power $P_s$=Source power | Battery Voltage($V_{dc}$) | Controller Logic 1:V2G 0:G2V |
|---|---|---|---|
| 1. | X≥0 | $V_{dc}$≥0.75($V_{rated}$) | 1 |
|  |  | 0.2($V_{rated}$)≤$V_{dc}$<0.75($V_{rated}$) | 0 |
|  |  | $V_{dc}$<0.2($V_{rated}$) | 0 |
| 2. | X<0 | $V_{dc}$≥0.75($V_{rated}$) | 1 |
|  |  | 0.2($V_{rated}$)≤$V_{dc}$<0.75($V_{rated}$) | 0 |
|  |  | $V_{dc}$<0.2($V_{rated}$) | 0 |

Fig. 3. Logic table for x and $V_b$

### D. State space modelling of plant

For the designing of MPC controller the plant is modelled by considering input as grid voltages i.e. $V_{ga}, V_{gb}, V_{gc}$ and output as inverter voltages $V_{ia}, V_{ib}, V_{ic}$. Here R,Y, and B phases are depicted as a,b, and c respectively.

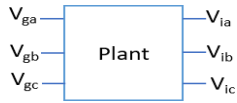

Fig. 4. Plant Model

The state space model of plant in continuous time domain is given as:

$$\dot{x}(t) = Ax(t) + Bu(t) \quad (1)$$

$$y(t) = Cx(t) + Du(t) + d(t); y(t) = \begin{bmatrix} V_{ia} \\ V_{ib} \\ V_{ic} \end{bmatrix}; u(t) = \begin{bmatrix} V_{ga} \\ V_{gb} \\ V_{gc} \end{bmatrix} \quad (2)$$

Here, matrices $x(t)$ is of dimension 18 × 1, $y(t)$ is of dimension 3 × 1, and $u(t)$ is of dimension 3 × 1. Matrix $A$ of dimension 18 × 18. Matrix $B$ is of dimension 18 × 3. Matrix $C$ is of dimension 3 × 18. It is implicitly assumed that the input $u(t)$ can't affect the output $y(t)$ simultaneously because of the control horizon control principle, which states that present plant knowledge is necessary for prediction and control. Hence, in the plant model, $D = 0$.

### E. Mathematical modelling of MPC for %THD optimization

The installation of a continuous-time predictive control system is done in a digital environment, despite the model and design being based on continuous time, which may benefit systems with a rapid or irregular sampling rate. To maximize the future behaviour of the plant output, $y(t)$, model predictive control's central design philosophy is to compute the trajectory of a future controlled variable, $u(t)$. A constrained window of time is used to complete the optimization. A window's length ($T_p$) and initial time ($t_i$) serve as the parameters for this time-dependent window for optimization. The window is taken from a starting time $t_i$ to a final time $t_i + T_p$. The window's size $T_p$ remains unchanged. The prediction horizon is the same as the discrete-time MPC in that it is equal to the size of the moving horizon window $T_p$ [24].

The plant model with $m$ inputs, $q$ outputs, $N_s$ number of states, $N_p$ number of prediction horizons and $N_c$ number of control horizons. which follows conditions as $m \geq q$. For this system, the values are given as $m = 3$, $n = 3$, $N_p = 10$, $N_c = 3$ and $T_s = 10 \mu sec$. Converting the continuous state space model of equation 1,2 into the discrete domain with sampling time $T_s$ is given by:

$$x((k+1)T_s) = G(T_s)x(kT_s) + H(T_s)u(kT_s) \quad (3)$$

$$y(kT_s) = Cx(kT_s) + d(kT_s) \quad (4)$$

$$G(T_s) = e^{AT_s}; H(T_s) = (e^{AT_s} - I)BA^{-1} \quad (5)$$

For simplicity, $kT_s$ is represented as $k$. The discrete modes are one step ahead prediction models, i.e. given data at sample $k$, one can determine data at sample $k + 1$ with the consideration of disturbances as unvarying i.e. $d(k+1) = d(k)$.

$$y(k+1) = CAx(k) + CBu(k) + d(k) \quad (6)$$

For n steps ahead prediction states are given as:

$$x(k+n) = G^n x(k) + G^{n-1} Hu(k) + G^{n-2} Hu(k+1) \\ + \ldots + GHu(k+n-2) + Hu(k+n-1) \quad (7)$$

$$y(k+n) = CG^n x(k) + CG^{n-1} Hu(k) + G_{n-2} Hu(k+1) \\ + \ldots + GHu(k+n-2) + Hu(k+n-1) + d(k) \quad (8)$$

$$\Delta x(k+1) = G\Delta x(k) + H\Delta u(k); \Delta y(k+1) = C\Delta x(k+1) \quad (9)$$

Choosing a new state variable vector $x(k) = [\Delta x(k)^T y(k)^T]$, we have:

$$\begin{bmatrix} \Delta x(k+1) \\ y(k+1) \end{bmatrix} = \overbrace{\begin{bmatrix} G_m & O_{3\times 18}^T \\ CG & I_{3\times 3} \end{bmatrix}}^{G} \begin{bmatrix} \Delta x(k) \\ y(k) \end{bmatrix} + \overbrace{\begin{bmatrix} H_m \\ CH \end{bmatrix}}^{H_m} \Delta u(k) \quad (10)$$

$$y(k) = \overbrace{[O_{3\times 18} \; \vdots \; I_{3\times 3}]}^{C_m} \begin{bmatrix} \Delta x(k+1) \\ y(k+1) \end{bmatrix} \quad (11)$$

For notation simplicity, equation 10 and 11, are written as:

$$x(k+1) = G_m x(k) + H_m \Delta u(k); \; y(k) = C_m x(k) \quad (12)$$

Here, $x_k$ is of order $21 \times 1$, $G_m$ is of order $21 \times 21$, $H_m$ is of order $21 \times 3$, $C_m$ is of order $3 \times 21$, and $y(k)$ is of order $3 \times 1$. For the MIMO system, $Y$ and $\Delta U$ for all phases together, is defined as:

$$\Delta U_{9\times 1} = \left[ \Delta u(k_i)^T \; \Delta u(k_i+1)^T \ldots \Delta u(k_i+N_c-1)^T \right]^T \quad (13)$$

Here, $\Delta u(k_i)$ is a column matrix with elements as $\Delta u_{ia}(k_i), \Delta u_{ib}(k_i)$, and $\Delta u_{ic}(k_i)$. As $N_c = 3$, So $\Delta U$ comes out to be a matrix of $9 \times 1$ order.

$$Y_{30\times 1} = \left[ y(k_i+1 \mid k_i)^T \; y(k_i+2 \mid k_i)^T \ldots y(k_i+N_p \mid k_i)^T \right]^T \quad (14)$$

Here, $y(k_i+1 \mid k_i)$ is a column matrix with elements as $y_a(k_i+1 \mid k_i), y_b(k_i+1 \mid k_i)$, and $y_c(k_i+1 \mid k_i)$. Based on state space model $G_m$, $H_m$ and $C_m$, the future state variables are calculated sequentially using the set of future control parameters:

$$x(k_i+N_p \mid k_i) = G_m^{N_p} x(k_i) + G_m^{N_p-1} H_m \Delta u(k_i) + G_m^{N_p-2} H_m \Delta u(k_i+1)$$
$$+ G_m^{N_p-N_c} H_m \Delta u(k_i+N_c-1).$$

Effectively, we have:

$$Y = F x(k_i) + \varphi \Delta U \quad (15)$$

$$F = \begin{bmatrix} CG_m \\ CG_m^2 \\ \vdots \\ CG_m^{N_p} \end{bmatrix}$$

$$\varphi = \begin{bmatrix} CH_m & 0 & \ldots & 0 \\ CG_m H_m & CH_m & \ldots & 0 \\ CG_m^2 H_m & CG_m H_m & \ldots & 0 \\ \vdots & & & \\ CG_m^{N_p-1} H_m & CG_m^{N_p-2} H_m & \ldots & CG_m^{N_p-N_c} H_m \end{bmatrix} \quad (16)$$

Assuming data vector that contains the set-point information is:

Here, $R_s$ is of order $30 \times 1$ and $\bar{R}_s$ is of order $30 \times 3$. The cost function $J$ for control objective is defined as:

$$J = (R_s - Y)^T (R_s - Y) + \Delta U^T \bar{R} \Delta U \quad (18)$$

Here, $\bar{R}$ is a block matrix with $m$ blocks defined as $r_w I_{N_c \times N_c}$ and has dimension equal to the dimensions of $\varphi^T \varphi$. For the desired closed loop performance $r_w$ employed as a tuning parameter. Without taking into account the size of control adjustments, the difference between projected $Y$ and $R_s$ is decreased i.e. $r_w = 0$, which implies $\bar{R} = 0$. To find the optimal value of $\Delta U$ that minimize $J$ [24], by using equation 15, $J$ is defined as:

$$J = (R_s - Fx(k_i))^T (R_s - Fx(k_i))$$
$$- 2\Delta U^T \Phi^T (R_s - Fx(k_i)) + \Delta U^T (\Phi^T \Phi) \Delta U \quad (19)$$

By finding first derivative of $J$ with respect to $\Delta U$ to calculate optimal value of $\Delta U$ for % THD optimization, is calculated as:

$$\frac{\partial J}{\partial \Delta U} = -2\Phi^T (R_s - Fx(k_i)) + 2(\Phi^T \Phi) \Delta U = 0 \quad (20)$$

The incremental optimal control within one optimization window is given by:

$$\Delta U = (\Phi^T \Phi)^{-1} (\Phi^T \bar{R}_s r(k_i) - \Phi^T F x(k_i)) \quad (21)$$

where, $\varphi^T \varphi$ has dimension $mN_c \times mN_c$ and $\varphi^T F$ has dimension $mN_c \times n$, and $\varphi^T \bar{R}$ equals the last $q$ columns of $\varphi^T F$, because the last column of $F$ is identical to $\bar{R}_s$. The set-point signal is $r(k_i) = [r_1(k_i) \; r_2(k_i) \; \ldots \; r_q(k_i)]^T$ as the $q$ set-point signals to the multi-output system at sample time $k_i$, within a prediction horizon. The goal of a predicted control system is to get the anticipated output as close as feasible to the set-point signal. Applying the receding horizon control principle, the first $m$ elements in $\Delta U$ are taken to form the incremental optimal control:

$$\Delta u(k_i) = [I_m \; \underbrace{o_m \; \ldots \; o_m}_{N_c}] (\Phi^T \Phi)^{-1} (\Phi^T \bar{R}_s r(k_i) - \Phi^T F x(k_i)) \quad (22)$$

where $I_m$ and $O_m$ are, respectively, identity and zero matrix with dimension $m \times m$.

### F. Model Predictive controller design

The designed state-space model from equations 1 and 2 are used to develop the predictive model controller using the MPC toolbox of MATLAB. In this, the output ($y_{3\times 1}$) is taken the same as measured output ($mo_{3\times 1}$), and input is taken as manipulating variable ($mv$) of all phases, i.e. $mv_a, mv_b$, and $mv_c$. The reference (ref) is taken as three sinusoidal signals

$$R^{-}{}_s = \begin{bmatrix} 1 & \cdots & 1 \\ 1 & \cdots & 1 \end{bmatrix}_T^{N_p} \; ; R_s = R^{-}{}_s r(k_i) = \begin{bmatrix} 1 & \cdots & 1 \\ 1 & \cdots & 1 \end{bmatrix}_T^{N_p} \begin{bmatrix} r_{ga}(k_i) \\ r_{gb}(k_i) \\ r_{gc}(k_i) \end{bmatrix}$$

(17)

with amplitude, phase and frequency same as grid voltage, which is done to make $J = 0$ so that %THD will be in the standard limit. The designed MPC block is used directly in the original model shown in Fig.1.

## III. RESULTS

This section shows results, depicting a comparative study of power quality factors, such as total harmonic distortion (THD)

in grid voltage, grid current, inverter output voltage, inverter output current, and grid frequency variation in closed loop configuration of G2V/V2G action with and without MPC.

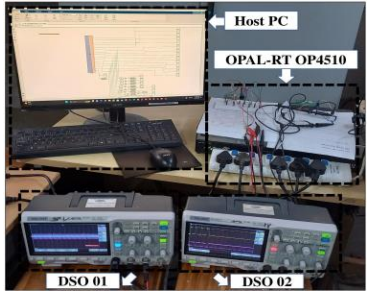

Fig. 5. Real Time Simulation Setup for Model Predictive Control (MPC) of G2V/V2G charging system using OPAL-RT OP4510

*A. Real-time digital environment result*

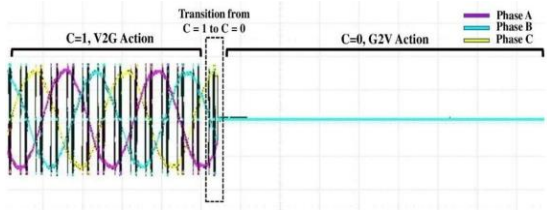

Fig. 6. Inverter Output Voltage in G2V/V2G mode without MPC

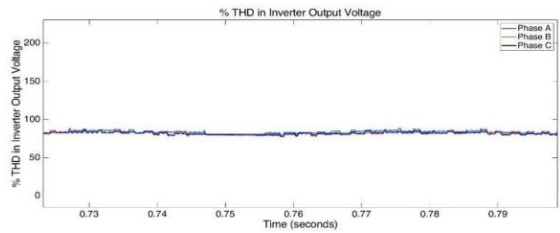

Fig. 7. % THD in Inverter Output Voltage during V2G mode without MPC

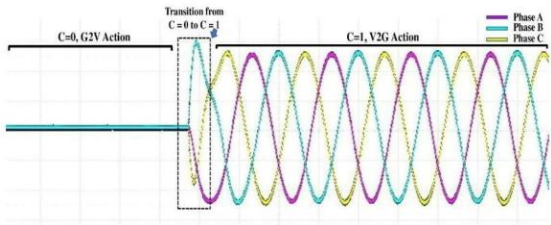

Fig. 8. Inverter Output Voltage in G2V/V2G mode with MPC

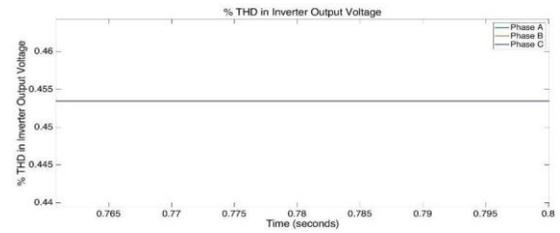

Fig. 9. % THD in Inverter Output Voltage during V2G mode with MPC

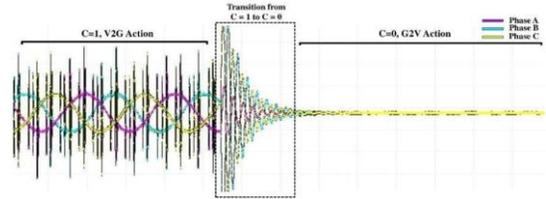

Fig. 10. Inverter Output current in G2V/V2G mode without MPC

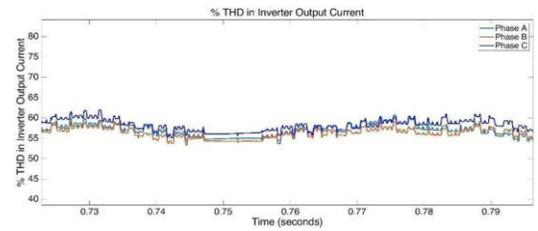

Fig. 11. % THD in Inverter Output current during V2G mode without MPC

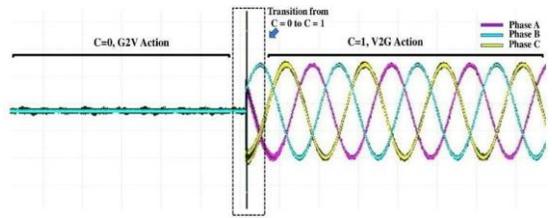

Fig. 12. Inverter Output current in G2V/V2G mode with MPC

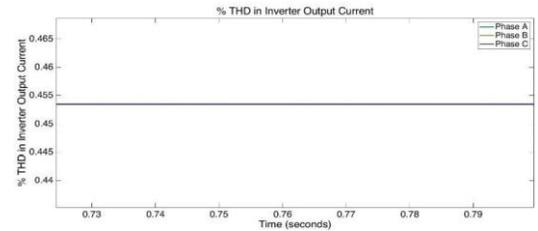

Fig. 13. % THD in Inverter Output current during V2G mode with MPC

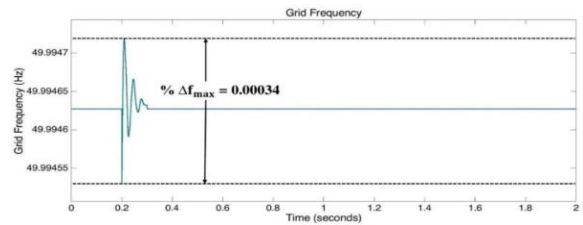

Fig. 14. Variation in Grid Frequency with MPC

TABLE I
COMPARISON OF % THD IN INVERTER PARAMETERS WITHOUT AND WITH MPC

| Sl. No. | Parameters | %THD without MPC | % THD with MPC | % Improvement |
|---|---|---|---|---|
| 1. | Grid Voltage | 60 - 70 | 0.45 | 99.31 |
| 2. | Grid Current | 85 - 95 | 0.45 | 99.43 |
| 3. | Inverter Output Voltage | 97 | 0.45 | 99.53 |
| 4. | Inverter Output Current | 55 - 60 | 0.45 | 99.17 |

Fig. 6 and Fig.8 shows Inverter Output Voltage in G2V/V2G mode without and with MPC, Fig.7 and Fig.9 shows its % THD during V2G mode without and with MPC.Fig. 10 and Fig.12 shows Inverter Output current in G2V/V2G mode without and with MPC, Fig.11 and Fig.13 shows its % THD during V2G mode without and with MPC. Thus we can clearly observe that without MPC, the % THD in all inverter parameters are violating the IEEE 519-2022 grid code standard but with MPC, they are satisfying the standards.

IV. CONCLUSION

An optimal closed-loop control of G2V/V2G action with and without a Model Predictive Controller (MPC) is investigated in this paper. The following conclusion is made based on the results as follows: (1)The hardware results have been achieved in MATLAB/Simulink and OPAL-RT (OP4510) real-time digital simulator environment under varying battery voltage, load power and source power conditions. (2)The proposed methodology shows the THD reduction to nearly 0.45% in Grid voltage, Grid current, inverter output voltage and inverter output current. (3)The proposed methodology shows the waveform quality improvement of grid voltage, grid current, inverter output voltage, and inverter output current to nearly 99%. (4)The system's frequency is maintained constant at 50 Hz with a very low deviation as compared to without MPC system.

REFERENCES

[1] S. Franzò and A. Nasca, "The environmental impact of electric vehicles: a comparative LCA-based evaluation framework and its application to the Italian context," 2020 Fifteenth International Conference on Ecological Vehicles and Renewable Energies (EVER), Monte-Carlo, Monaco, 2020, pp. 1-4, doi: 10.1109/EVER48776.2020.9243006
[2] S. V. P. Singh and P. Agnihotri, "ANN Based Modelling of Optimal Passive Cell Balancing," 2022 22nd National Power Systems Conference (NPSC), New Delhi, India, 2022, pp. 326-331, doi: 10.1109/NPSC57038.2022.10069440.
[3] Q. Cao, J. Zhang, H.G. Tian. Model Predictive Control of Three-Phase Rectifier for Electric Vehicle Charging. In 2021 IEEE International Conference on Predictive Control of Electrical Drives and Power Electronics (PRECEDE) 2021 Nov 20 (pp. 915-920). IEEE.
[4] T. He, M. Wu, D. D. C. Lu, R. P. Aguilera, J. Zhang, and J. Zhu, (2019). Designed dynamic reference with model predictive control for bidirectional EV chargers. IEEE Access, 7, 129362-129375.
[5] M. Parvez,, Saad Mekhilef, Nadia ML Tan, and Hirofumi Akagi. "Model predictive control of a bidirectional AC-DC converter for V2G and G2V applications in electric vehicle battery charger." In 2014 IEEE Transportation Electrification Conference and Expo (ITEC), pp. 1-6. IEEE, 2014.
[6] Joerg Dannehl, Christian Wessels, and Friedrich Wilhelm Fuchs. "Limitations of voltage-oriented PI current control of grid-connected PWM rectifiers with LCL filters." IEEE transactions on industrial electronics 56, no. 2 (2008): 380-388.
[7] Riad Kadri, Jean-Paul Gaubert, and Gerard Champenois. "An improved maximum power point tracking for photovoltaic grid-connected inverter based on voltage-oriented control." IEEE transactions on industrial electronics 58, no. 1 (2010): 66-75.
[8] Yongchang Zhang, Jihao Gao, and Changqi Qu. "Relationship between two direct power control methods for PWM rectifiers under unbalanced network." IEEE Transactions on Power Electronics 32, no. 5 (2016): 4084-4094..
[9] Manaswi Srivastava, Jitendra Kumar Nama, and Arun Kumar Verma. "An efficient topology for electric vehicle battery charging." In 2017 IEEE PES Asia-Pacific Power and Energy Engineering Conference (APPEEC), pp. 1-6. IEEE, 2017.
[10] Noguchi, T., Tomiki, H., Kondo, S. and Takahashi, I., 1998. Direct power control of PWM converter without power-source voltage sensors. IEEE transactions on industry applications, 34(3), pp.473-479.
[11] T. He, D. Lu, and L. Li, "Model-predictive sliding-mode control for three phase AC/DC converters," IEEE Trans. Power Electron., vol. 33, no. 10, pp. 8982-8993, Oct. 2108. doi: 10.1109/TPEL.2017.2783859.
[12] J. A. Rohten et al., "Model Predictive Control for Power Converters in a Distorted Three-Phase Power Supply," IEEE Transactions on Industrial Electronics, vol. 63, no. 9, pp. 5838-5848, Sept. 2016.
[13] S. Vazquez, J. Rodriguez, M. Rivera, L. G. Franquelo and M. Norambuena, "Model Predictive Control for Power Converters and Drives: Advances and Trends," IEEE Transactions on Industrial Electronics, vol. 64, no. 2, pp. 935-947, Feb. 2017.
[14] Mohammadi, Fazel, Gholam-Abbas Nazri, and Mehrdad Saif. "A bidirectional power charging control strategy for plug-in hybrid electric vehicles." Sustainability 11, no. 16 (2019): 4317.
[15] De Luca, Felice, Vito Calderaro, and Vincenzo Galdi. "A fuzzy logic-based control algorithm for the recharge/v2g of a nine-phase integrated on-board battery charger." Electronics 9, no. 6 (2020): 946.
[16] Li, Yao, Liying Li, Chao Peng, and Jianxiao Zou. "An MPC based optimized control approach for EV-based voltage regulation in distribution grid." Electric Power Systems Research 172 (2019): 152-160.
[17] He, Tingting, Jianguo Zhu, Jianwei Zhang, and Linfeng Zheng. "An optimal charging/discharging strategy for smart electrical car parks." Chinese Journal of Electrical Engineering 4, no. 2 (2018): 28-35.
[18] Patil, Vinayak, and M. R. Sindhu. "An intelligent control strategy for vehicle-to-grid and grid-to-vehicle energy transfer." In IOP Conference Series: Materials Science and Engineering, vol. 561, no. 1, p. 012123. IOP Publishing, 2019.
[19] He, Tingting, Dylan Dah-Chuan Lu, Mingli Wu, Qinyao Yang, Teng Li, and Qiujiang Liu. "Four-quadrant operations of bidirectional chargers for electric vehicles in smart car parks: G2v, v2g, and v4g." Energies 14, no. 1 (2020): 181.
[20] Nisha, K. S., and Dattatraya N. Gaonkar. "Model predictive controlled three-level bidirectional converter with voltage balancing capability for setting up EV fast charging stations in bipolar DC microgrid." Electrical Engineering 104, no. 4 (2022): 2653-2665.
[21] Tan, A. S. T., Ishak, D., Mohd-Mokhtar, R., Lee, S. S., and Idris, N. R. N. (2018). Predictive control of plug-in electric vehicle chargers with photovoltaic integration. Journal of Modern Power Systems and Clean Energy, 6(6), 1264-1276.
[22] Pan, L., and Zhang, C. (2017). Model predictive control of a single-phase PWM rectifier for electric vehicle charger. Energy Procedia, 105, 4027-4033.
[23] Gil-González, W., Serra, F., Dominguez, J., Campillo, J., and Montoya, O. (2020, October). Predictive power control for electric vehicle charging applications. In 2020 IEEE ANDESCON (pp. 1-6). IEEE.
[24] L. Wang, Model Predictive Control System Design and Implementation using MATLAB®, Springer, London, 2009.